\documentclass{PoS}
\usepackage{epsfig}
\usepackage{subfigure}

\title{Measurements of dilepton continuum at the PHENIX experiment at RHIC}

\ShortTitle{Measurements of dilepton continuum at the PHENIX experiment at RHIC}

\author{\speaker{Alberica Toia}%
         \thanks{A footnote may follow.}\\
        SUNY, Stony Brook\\
        E-mail: \email{alberica@skipper.physics.sunysb.edu}}

\abstract{
PHENIX has measured the dielectron continuum in $\sqrt{s_{NN}}$=200 GeV Au+Au and p+p collisions.
In minimum bias Au+Au collisions the dielectron yield in the mass range between 150 and 750 MeV/c$^2$ is enhanced by a factor of
3.4$\pm$0.2(stat.)$\pm$1.3(syst.)$\pm$0.7(model) compared to the expectation from our model of hadron decays that well reproduces the mass spectrum in p+p collisions. 
The integrated yield increases faster with the centrality of the collisions than the number of participating nucleons, suggesting emission from scattering processes in the dense medium.
The continuum yield between the masses of the $\phi$ and the $J/\psi$ meson is consistent with expectations from correlated $c\bar{c}$ production, though other mechanisms are not ruled out.
}

\FullConference{Critical Point and Onset of Deconfinement
          4th International Workshop\\
		 July 9-13 2007\\
		 GSI Darmstadt,Germany}

\begin{document}
\section{Introduction}
Electron-positron pairs, or dileptons in general, have proven to be an 
excellent tool to study collisions of heavy ions at ultra-relativistic 
energies. Because leptons do not interact strongly, emission of dileptons 
from the hot matter created at RHIC should leave an imprint on the observed 
dilepton distributions. Emission from the hot matter may include thermal 
radiation and in-medium decays of mesons with short lifetimes, like the 
$\rho$ meson, while their spectral functions may be strongly modified. 
However, below the mass of the $\phi$ meson, these sources compete with a 
large contribution of e$^+$e$^-$--pairs from Dalitz decays of pseudoscalar 
mesons ($\pi^0, \eta, \eta'$) and decays of vector mesons ($\rho, \omega, 
\phi$). Above the $\phi$ meson mass up to 4.5 GeV/c$^2$, competing sources 
are dilepton decays of charmonia ($J/\psi, \psi'$) and semileptonic decays of 
$D$ and $\bar{D}$ mesons, correlated through flavor conservation, which lead 
to a continuum of masses. In addition to thermal radiation, energy loss of 
charm quarks in the medium might modify the continuum yield in this mass 
region.

%%%%%%%%%%%%%%%%%%%%%%%%%%%%%%%%%%%---previous measurements

The discovery of a large enhancement of the dilepton yield at masses below 
the $\phi$ meson mass in ion-ion collisions at the CERN SPS \cite{CER1} has 
triggered a broad theoretical investigation of modifications of properties of 
hadrons in a dense medium and of how these modifications relate to chiral 
symmetry restoration \cite{theory1}. These theoretical studies advanced 
from the availability of more precise data from CERN \cite{NA60_rho,CER2} and 
GSI \cite{HADES}. An enhanced yield was also observed at higher masses, above 
the $\phi$ meson mass \cite{NA50}. Preliminary NA60 data suggest that the 
enhancement can not be attributed to decays of D-mesons but may result from 
prompt production, as expected for thermal radiation \cite{NA60_therm}.

\section{Analysis}

%%%%%%%%%%%%%%%%%%%%%%%%%%%%%%%%%%%%---the phenix experiment
The PHENIX experiment at the Relativistic Heavy Ion Collider (RHIC) extends 
these measurements in a new energy regime by exploring Au+Au collisions at a 
center of mass energy of $\sqrt{s_{NN}}$=200 GeV. In this paper we present 
results from minimum bias data taken in 2004. Collisions were triggered 
and selected by centrality using beam-beam counters (BBC) and zero degree 
calorimeters (ZDC). The minimum bias trigger corresponds to $92^{+2.5}_{-3.0}$\% of the Au+Au inelastic cross-ection. We analyzed a sample of 8$\times 10^8$ minimum bias 
events.

%%%%%%%%%%%%%%%%%%%%%%%%%%%%%%%%%%%%%%---tracking and momentum resolution 

Electrons and positrons are reconstructed in the two central arm 
spectrometers of PHENIX \cite{NIM} using Drift Chambers (DC) and Pad Chambers (PC), located outside 
an axial magnetic field, which measure their momenta with an accuracy of 
$\sigma_p/p=0.7\%\oplus 1\%p/(\mathrm{GeV/c})$.

%%%%%%%%%%%%%%%%%%%%%%%%%%%%%%%%%%%%%%---eID (cut out and reference)
\subsection{Electron Identification}
Electrons are identified by hits in the Ring Imaging Cherenkov detector (RICH) 
and by matching the momentum with the energy measured in an electromagnetic 
calorimeter (EMCal) \cite{ppg066}. Below the Cherenkov threshold for pions ($p_T \leq 5 GeV/c$) electron misidentification due to random coincidence between hadron tracks and hits in the RICH leads to $\sim$20\% contamination.

%%%%%%%%%%%%%%%%%%%%%%%%%%%%%%%%%%%%%%--- acceptance

Each central arm covers $\left|\Delta\eta\right|\le0.35$ in pseudorapidity 
and $\pi/2$ in azimuthal angle. Because charged particles are deflected in 
the azimuthal direction by the magnetic field, the acceptance depends on the 
momentum and the charge of the particle, and also on the radial location of 
the detector component (DC, EMCal and RICH). The acceptance for a track with 
charge $q$, transverse momentum $p_{\rm T}$ and azimuthal emission angle 
$\phi$ can be described by:
\begin{equation} \label{eq:acc}
 \phi_{\rm min} \leq \phi+ q \frac{k_{\rm DC} (k_{RICH})}{p_{\rm T}} \leq \phi_{\rm max}
\end{equation}
where $k_{\rm DC}$ and $k_{\rm RICH}$ represent the effective azimuthal bend 
to DC and RICH ($k_{DC}=0.206$ rad GeV/c and $k_{RICH}=0.309$ rad GeV/c). One 
arm covers the region from $\phi_{\rm min}=\frac{-3}{16}\pi$ to $\phi_{\rm 
max}=\frac{5}{16}\pi$, the other arm from $\phi_{min}=\frac{11}{16}\pi$ to 
$\phi_{max}=\frac{19}{16}\pi$. Only electrons with $p_{\rm T}\ge$~200~MeV/c 
are used in the analysis. The photon conversion probability was minimized by 
installing a helium bag between the beam pipe and the DC, reducing the 
material to $\sim$0.4\% of a radiation length.

%%%%%%%%%%%%%%%%%%%%%%%%%%%%%%%%%%%%%%---all pairs
\subsection{Pair Analysis}
In an event the source of any electron or positron is unknown and therefore 
all electrons and positrons are combined into pairs, like-sign and unlike-sign.
%%%%%%%%%%%%%%%%%%%%%%%%%%%%%%%%%%%%%%---combinatorial background
This results in a large combinatorial background which must be removed. 
In experiments with equal acceptance for electrons and positrons, the background can be measured directly through the geometrical mean of the like sign pairs $2\sqrt{N_{++}N_{--}}$. In PHENIX the different acceptance (see Equation \ref{eq:acc}) prohibits this approach and the combinatorial background must be determined using mixed events. While the shape of the like and unlike sign background are not the same, the normalization of the mixed events unlike sign background is still given by $2\sqrt{N_{++}N_{--}}$. Also the shape of the like sign background can be compared to the real data to determine how well the background shape is reproduced in the event mixing.
The background is computed with a mixed event technique, which combines tracks 
from different events with similar topology (centrality, collision 
vertex, reaction plane).

%%%%%%%%%%%%%%%%%%%%%%%%%%%%%%%%%%%%%%%---remove background pairs: 
\subsubsection{Combinatorial Background from Mixed Events}
The mixed events background must have the same shape as the background in real events.
In order to achieve this, all unphysical correlations that 
arise from overlapping tracks or hits in the detectors, mostly in the RICH, 
must be eliminated, because they can not be reproduced by mixed events. If 
hits of both tracks of a pair overlap in any detector, the event is rejected. 
While the fraction of events removed by this rejection is $\sim$ 0.08\%, the fraction of pairs removed varies from 4\% in the most central to 2\% in the most peripheral collisions. 
%%%%%%%%%%%%%%%%%%%%%%%%%%%%%%%%%%%%%%---shape
\begin{figure}[!h]
\mbox{
\parbox{8cm}{\subfigure[before event cut]{\epsfig{file=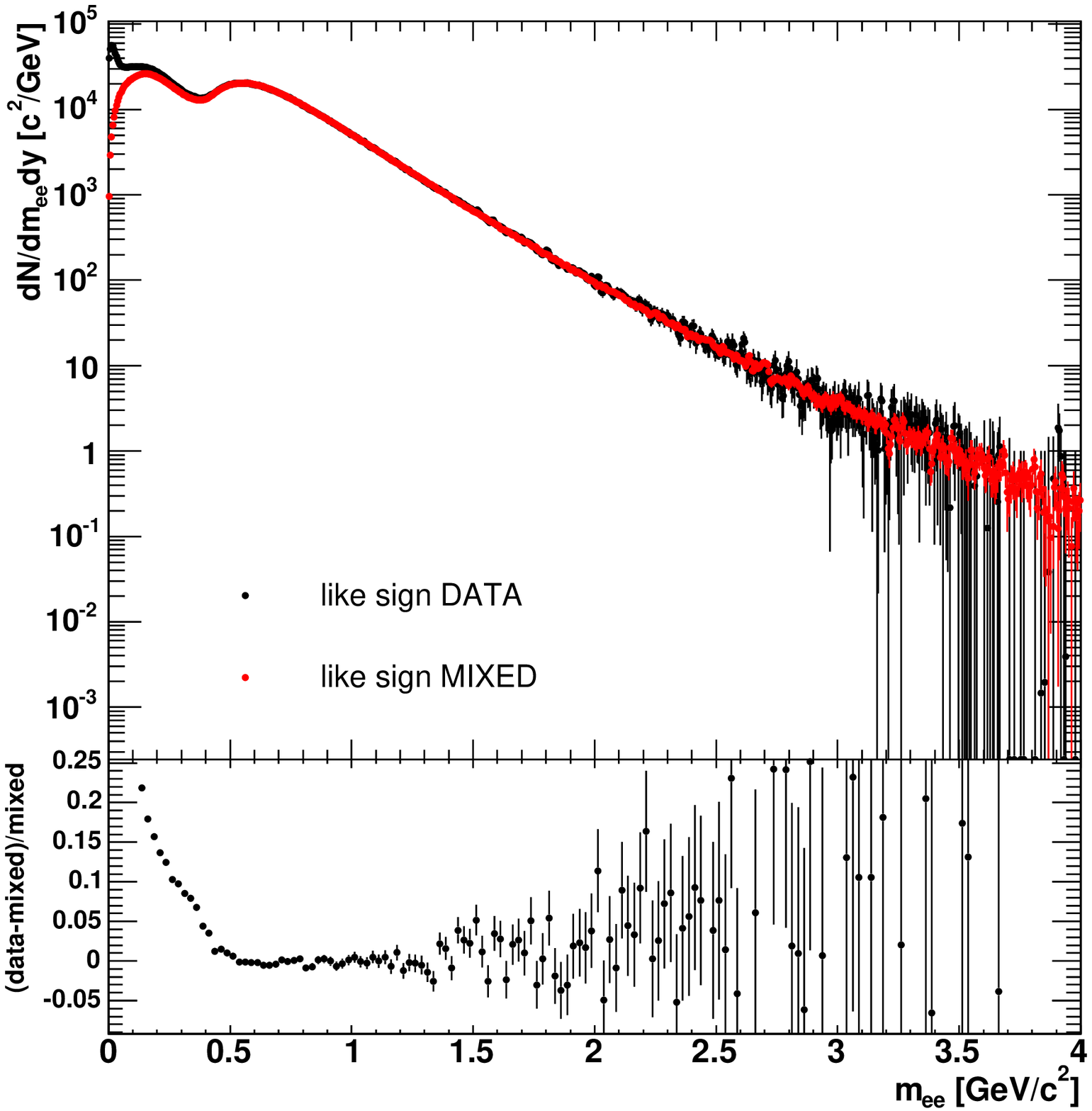,width=0.5\textwidth}}}
\parbox{8cm}{\subfigure[after event cut]{\epsfig{file=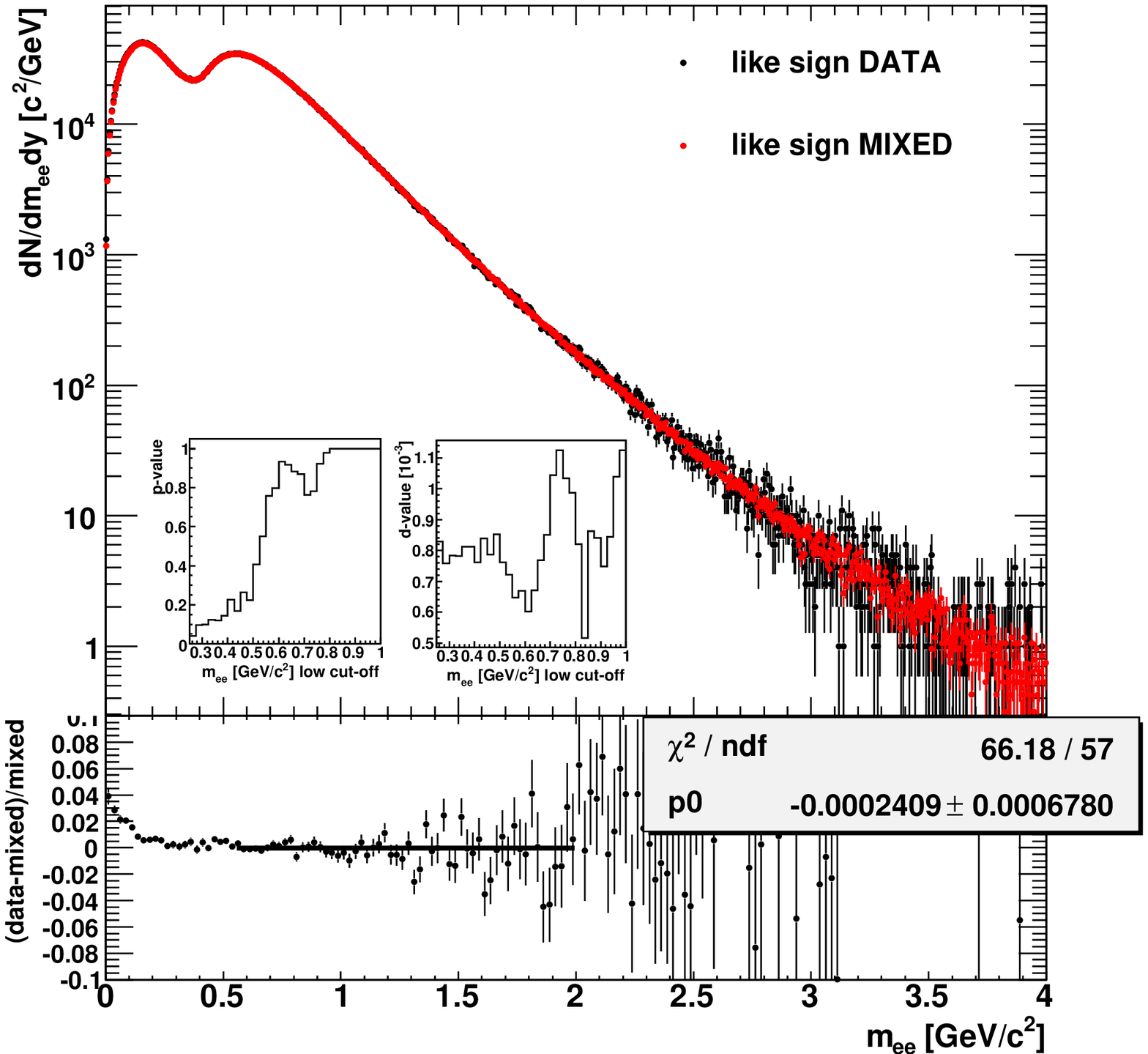,width=0.5\textwidth}}}}
 \caption{\label{fig:like}Like-sign distribution for real and mixed events, before and after the event cut that removes the unphysical correlations from overlapping tracks. The bottom panels show the ratio of $(data-mixed)/mixed$.}
\end{figure}
Comparing measured like-sign pairs with the mixed combinatorial background 
shows that the mixing technique reproduces the shape within the statistical 
accuracy of the data. Fig.~\ref{fig:like} shows the like-sign distribution for real and mixed events: before the rejection of overlapping tracks the mixed events distribution does not reproduce the shape of the measured like sign distribution.
After removing events with overlapping tracks, the like sign distributions from real and mixed events are in a very good agreement, with no significant trend deviating from a constant, as shown in the right panel of Figure \ref{fig:like}. A small signal from correlated background remains at low masses, see below. To compare the shape of the data to the mixed events we calculate the ratio $(data-mixed)/mixed$ and fit it with a constant above the $\eta$ mass (550 MeV/c$^2$). The result is (- 0.000241$\pm$0.000678) with a $\chi^2$/n.d.f. of 1.14.
The value of the test statistic $\chi^{2}$ performed on the real and mixed events is equal to 0.509 with a p-value equal greater than 0.999.
The Kolmogorv-Smirnov test gives a maximum deviation (see inlet of Fig.~\ref{fig:like}) of 0.1\% for any centrality bin or mass range which is small compared to the uncertainty of the absolute normalization of the mixed event background. The corresponding p-value therefore confirms that the hypothesis of compatibility of the two distributions can be accepted for any commonly used significant level.
%%%%%%%%%%%%%%%%%%%%%%%%%%%%%%%%%%%%%%---2sqrt

The absolute normalization of the unlike-sign combinatorial background is 
given by the geometrical mean of the observed positive and negative like-sign 
pairs $2\sqrt{N_{--}N_{++}}$, where, in principle, $N_{--}$ and $N_{++}$ are 
the measured number of like-sign pairs.
%%%%%%%%%%%%%%%%%%%%%%%%%%%%%%%%%%%%%%---like sign signal
There is a small correlated signal also in the observed like-sign pairs, 
which occurs if there are two e$^+$e$^-$--pairs in the final state of a 
meson, e.g. double Dalitz decays, Dalitz decays followed by a conversion of 
the decay photon or two photon decays followed by conversion of both photons. 
These ``cross'' pairs have small masses, typically less than the $\eta$ mass 
(550 MeV/c$^2$).

%%%%%%%%%%%%%%%%%%%%%%%%%%%%%%%%%%%%%%--- normalization to the like sign

We therefore determine $N_{--}$ and $N_{++}$ by integrating the mixed event 
distributions after they were normalized to the $7.5 \times 10^6$ like-sign 
pairs measured above 700 MeV/c$^2$. $N_{--}$ and $N_{++}$ are determined with 
a statistical accuracy of 0.12\%.
%%%%%%%%%%%%%%%%%%%%%%%%%%%%%%%%%%%%%%---kappa correction
The normalization factor is corrected by 1.004 $\pm$ 0.002 because the
event rejection removes 10$\pm$5\% more like-sign than unlike-sign pairs. We note that the error, estimated with mixed events, is a conservative upper limit.
While the pair loss depends on the centrality and varies from 85\% in the most peripheral to 96\% in the most central, the ratio of like to unlike sign pairs lost was found to be independent of centrality.

%%%%%%%%%%%%%%%%%%%%%%%%%%%%%%%%%%%%%%---sys
Adding the statistical error and the uncertainty due to the event rejection 
in quadrature gives an accuracy of 0.25\% on the normalization, which is conservative and represents an upper limit for the systematic uncertainty.

%%%%%%%%%%%%%%%%%%%%%%%%%%%%%%%%%%%%%%---photon conversion
\subsubsection{Photon Conversion}
After subtraction of the combinatorial background, physical background from 
photon conversions and cross pairs is removed. Since the tracking assumes 
that the e$^+$e$^-$--pair originates at the collision vertex, pairs from 
photons that convert in or outside of the beampipe (off-vertex) are reconstructed with 
an artificial opening angle which leads to an invariant mass that is proportional to the radius at which the conversion occurs.
The opening angle though is oriented perpendicular to the magnetic field and a cut on the orientation of the opening angle in the field removes more than 98\% of the conversion pairs.
Fig.~\ref{fig:corrbg} (right panel) shows the signal after subtraction of the combinatorial background and (dashed histogram) the contribution from the conversion pairs. The spectrum of those pairs represent a ``tomography'' of the material in the spectrometer: the flat background from air conversion extends up to $\sim$ 300 MeV/c$^2$, i.e. the entrance window of the DC after which the electrons do not bend anymore because the region is field-free, and are therefore removed with a high p$_T$ cut, while the peaks correspond to the beampipe (r=4 cm, or $m_{ee}$=20 MeV/c$^2$) and detector support structures (r$\sim$ 25 cm, $m_{ee}$=125 MeV/c$^2$).    
The full circles show the signal after subtraction of the conversion contribution.

%%%%%%%%%%%%%%%%%%%%%%%%%%%%%%%%%%%%%%---cross   
\subsubsection{Correlated Background}
Whenever there are two e$^+$e$^-$--pairs in the final state of a meson, cross pairs occur at the same rate as like and unlike-sign pairs; however the PHENIX acceptance is different for like and unlike sign pairs.
Because their rate is proportional to the conversion probability only, the ratio of cross pairs to cocktail (or to signal) is independent of centrality.
We performed a full scale Monte Carlo simulation of 10 million $\pi^{0}$ decays, flat in a rapidity range $-0.6\leq y \leq 0.6$, an azimuthal range $0\leq \phi \leq 2\pi$ and flat in p$_{T}$ weighted according to \cite{pi0,pich}. 
The simulations show that the rate of unlike-sign cross pairs accepted in PHENIX is 44\% of the 
rate for like-sign cross pairs. With a fast Monte Carlo, we simulated the like and unlike cross contribution from $\eta$'s, which was added to the $\pi^{0}$, scaled by the known $\eta$/$\pi^{0}$ branching ratio and the same relative acceptance of 44\%.
\begin{figure}[!h]
\mbox{
\parbox{7.5cm}{\subfigure[conversion]{\epsfig{file=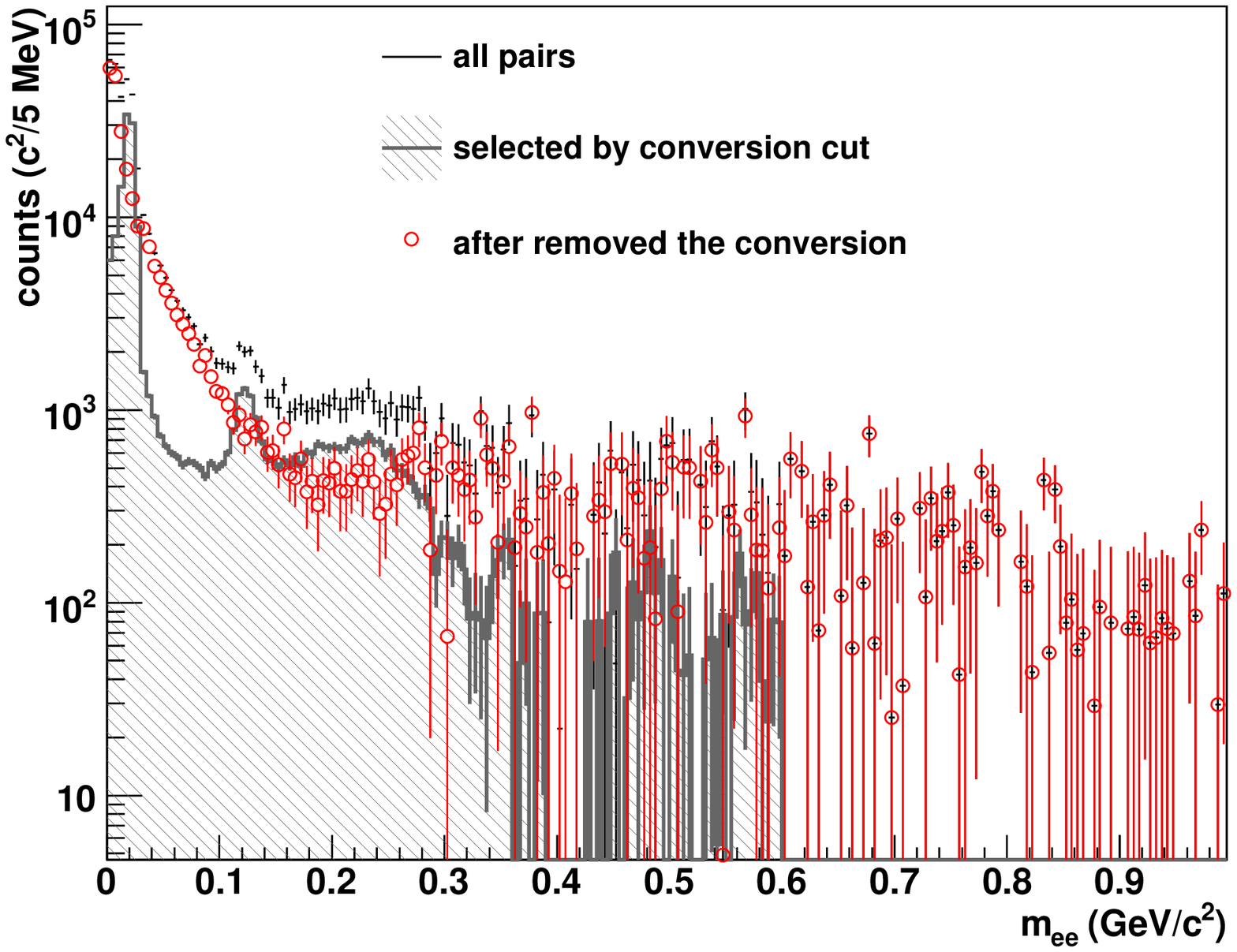,width=0.5\textwidth}}}
\parbox{8.5cm}{\subfigure[cross]{\epsfig{file=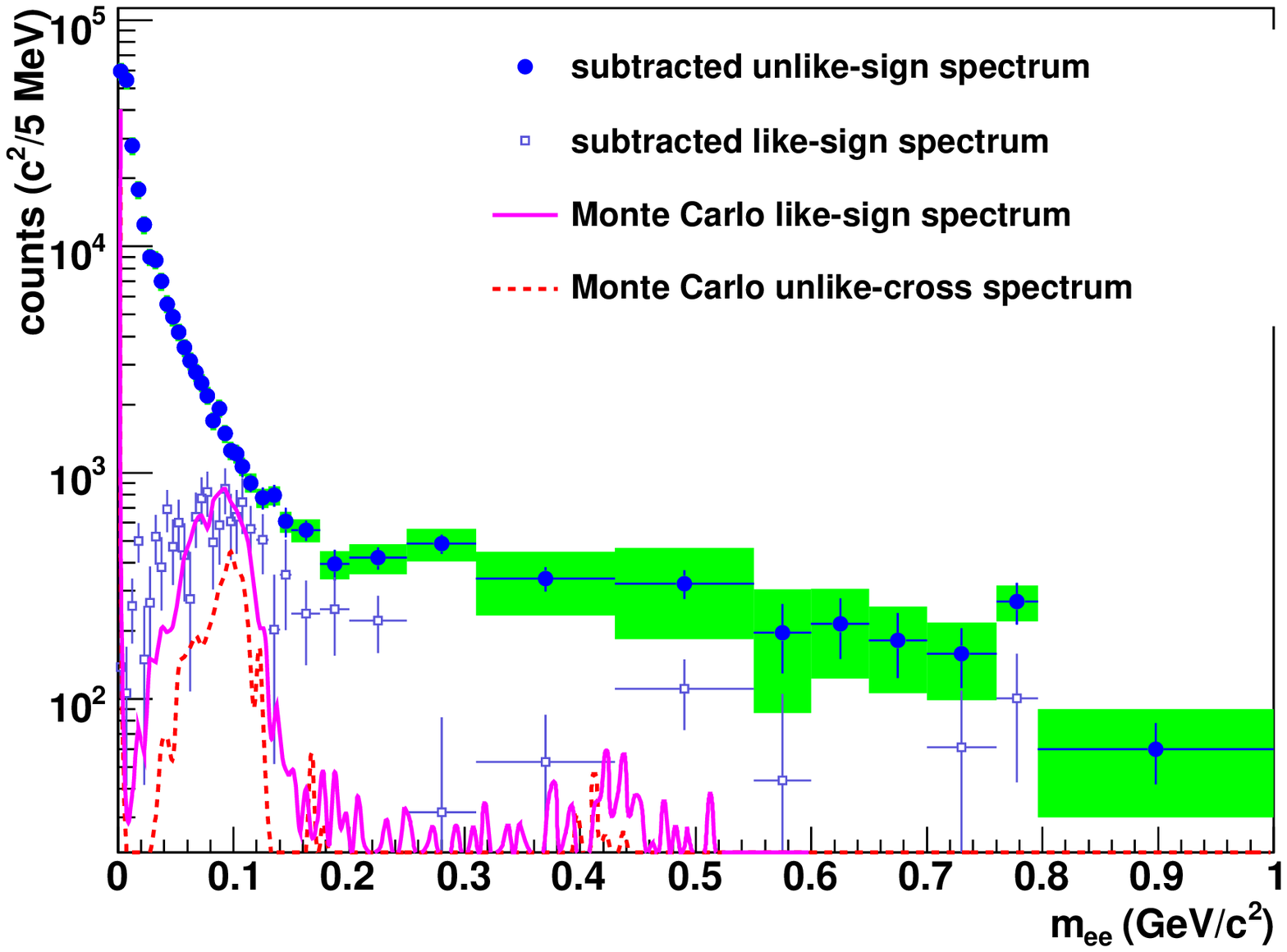,width=0.5\textwidth}}}}
 \caption{\label{fig:corrbg}Correlated background from conversion pairs (left panel) and cross pairs (right panel)}
\end{figure}

Fig.~\ref{fig:corrbg} (right panel) demonstrates how we subtracted this contribution.
To determine the rate of unlike-sign cross pairs to be subtracted from the signal, we scale the simulated like-sign cross pair distribution (full curve) to the observed like-sign signal (open squares), obtained by subtraction of the mixed event 
background normalized above 700 MeV/c$^2$. 
We note that the like-sign signal is well described by the Monte Carlo simulation up to 150 MeV/c$^2$, but not in the region $150\leq m_{ee}\leq 250 MeV/c^2$. 
The simulated unlike-sign cross pair distribution (dashed curve) is scaled by the same factor and subtracted from the 
unlike-sign signal (full circles). In the region $150 \leq m_{ee}\leq 250 MeV/c^2$ we subtract either the unlike-sign cross pair distribution calculated with the Monte Carlo (full curve) or -point by point- 44\% of the measured like-sign yield (open squares) and we take the difference of these two subtractions as systematic uncertainty which is $\leq$ 9\% of the final yield.

\section{Results}
%%%%%%%%%%%%%%%%%%%%%%%%%%%%%%%%%%%%%%---RESULTS
%%%%%%%%%%%%%%%%%%%%%%%%%%%%%%%%%%%%%%---fig 1

Figure~\ref{fig:mass}\cite{ppg075} shows the mass distribution of e$^+$e$^-$--pairs, the 
normalized mixed event background ({\it{B}}), and the signal yield ({\it{S}}) 
obtained by subtracting the mixed event background, the cross pairs and the 
conversion pairs. The insert shows the signal-to-background ratio ($S/B$). 
The systematic errors (boxes) reflect the error from the background 
subtraction, which is given by $\delta_S/S~=~0.25\%~\cdot~B/S$, added in 
quadrature to the uncertainty due to the cross pair subtraction, assumed to 
be $9\%S$ below 600 MeV/c$^2$. Despite the small $S/B$ ratio, the vector 
meson resonances $\omega$, $\phi$ and $J/\psi$ which decay directly to 
e$^+$e$^-$, and an e$^+$e$^-$--pair continuum are visible up to 4.5 GeV/c$^2$.

%%%%%%%%%%%%%%%%%%%%%%%%%%%%%%%%%%%%%%---converter
\subsection{Analysis of Runs with Increasing Conversion Material}

\begin{figure*}[t]
 \includegraphics[width=0.625\linewidth]{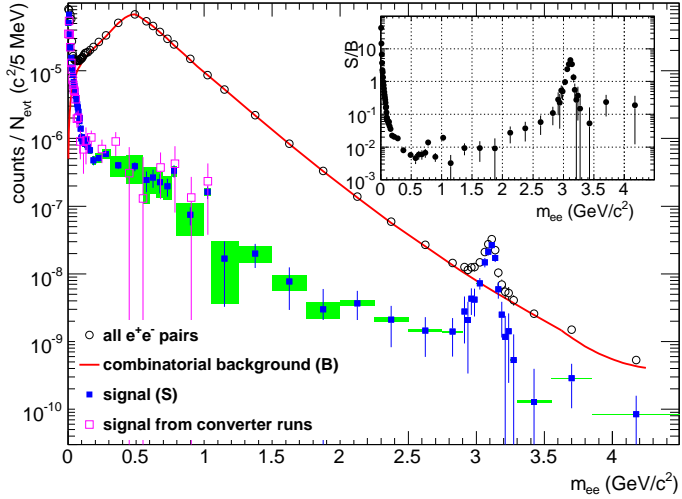}  
 \caption{\label{fig:mass}
Uncorrected mass spectra of all e$^+$e$^-$ pairs, mixed events background 
($B$) and signal ($S$) with statistical (bars) and systematic (boxes) 
uncertainties shown separately. The signal from the runs with additional 
converter is shown with statistical errors only. The insert shows the $S/B$ 
ratio. The mass range covered by each data point is given by horizontal bars.
}
\end{figure*}

In order to check the background subtraction, a subset of data (5$\times 
10^7$ events), taken with additional material wrapped around the beam pipe to 
increase the number of photon conversions \cite{ppg066}, was analyzed. Because the additional conversion leads to an increased electron multiplicity, in 
this data set the combinatorial background and the cross pair contribution is 
larger by a factor of $\sim$2.5. Thus the background would be much larger if there was an error in the background normalization.
\begin{figure}[!h]
\mbox{
\parbox{5cm}{\subfigure[70--300 MeV]{\epsfig{file=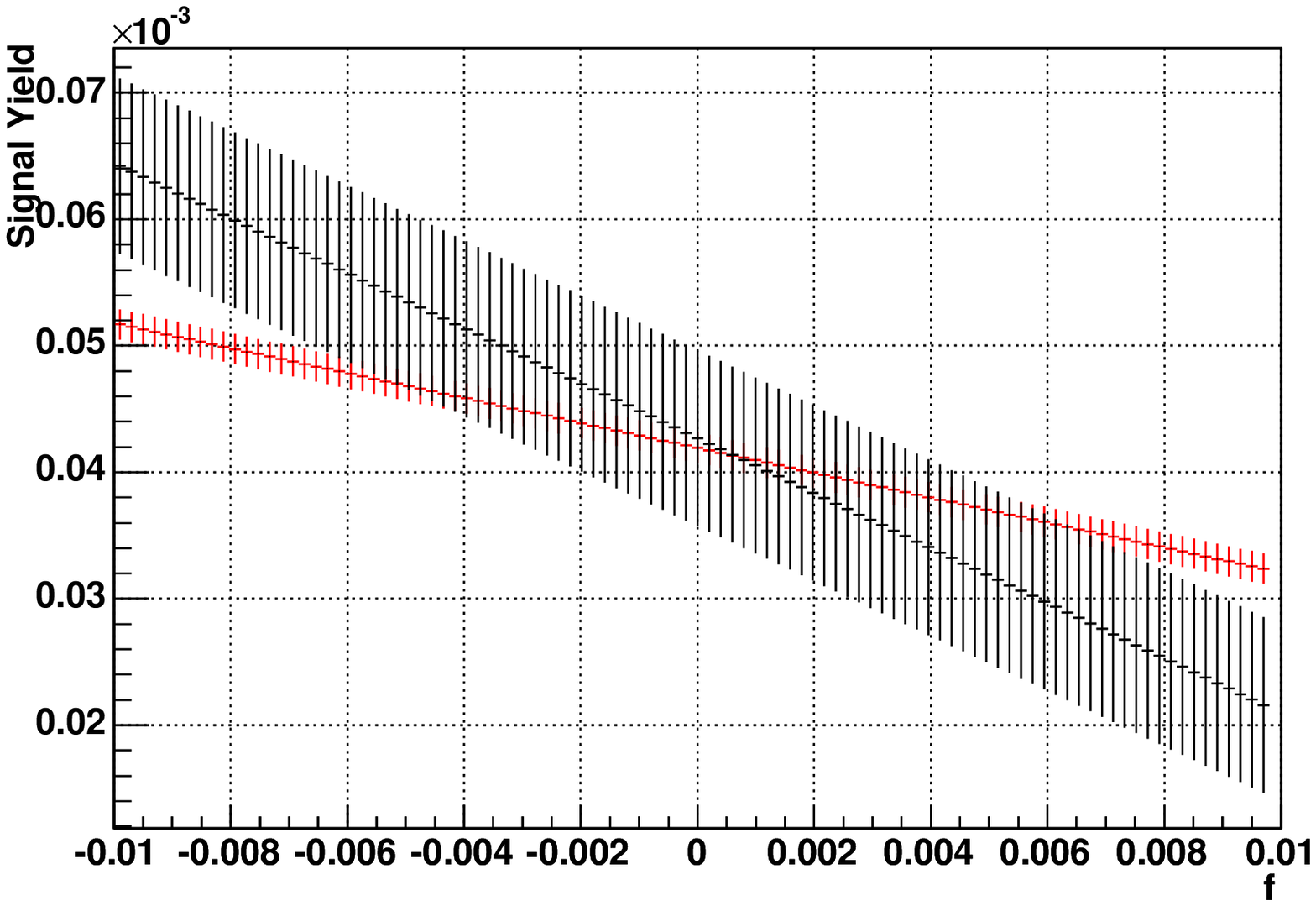,width=0.33\textwidth}}}
\parbox{5cm}{\subfigure[300--650 MeV]{\epsfig{file=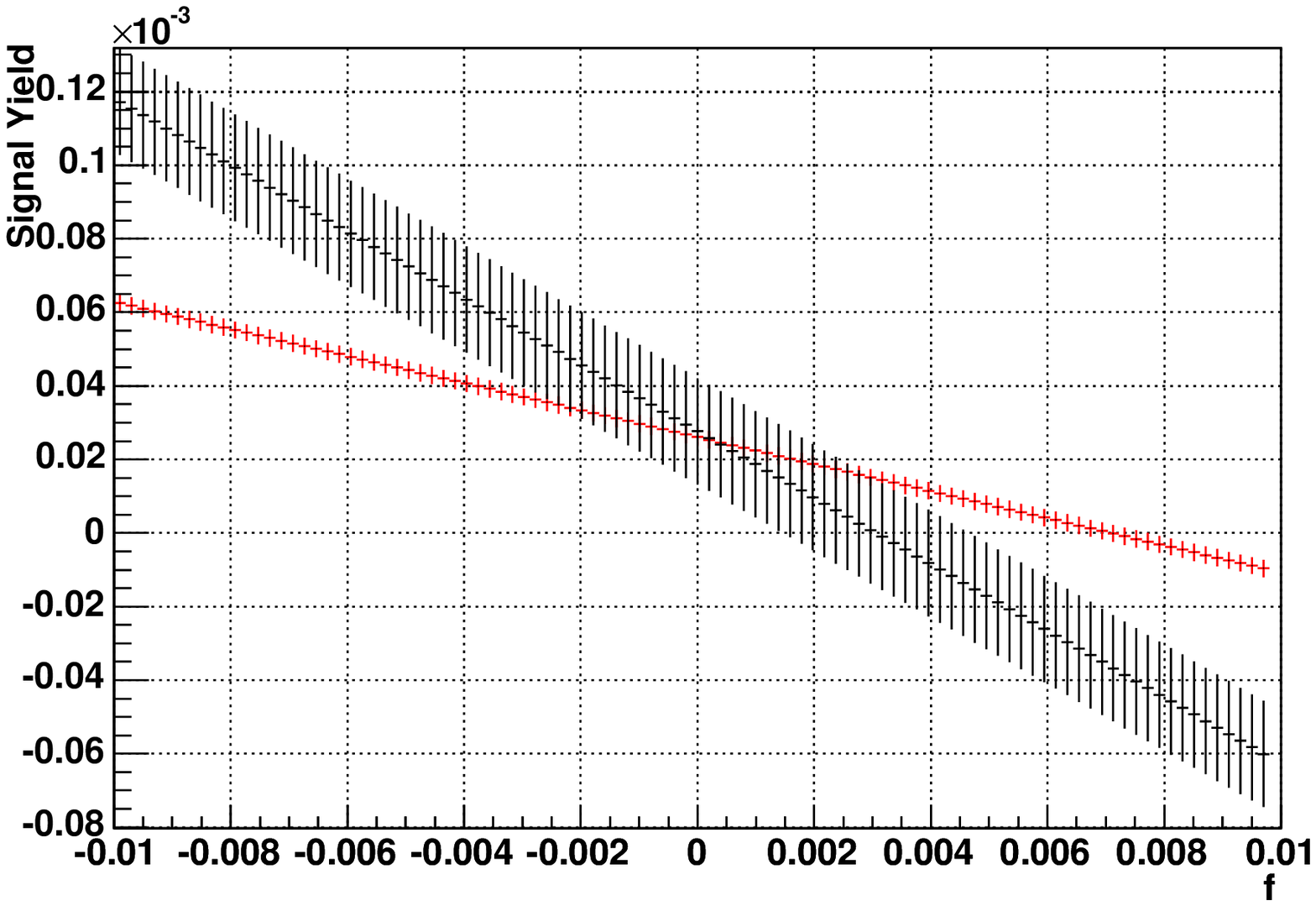,width=0.33\textwidth}}}
\parbox{5cm}{\subfigure[500--750 MeV]{\epsfig{file=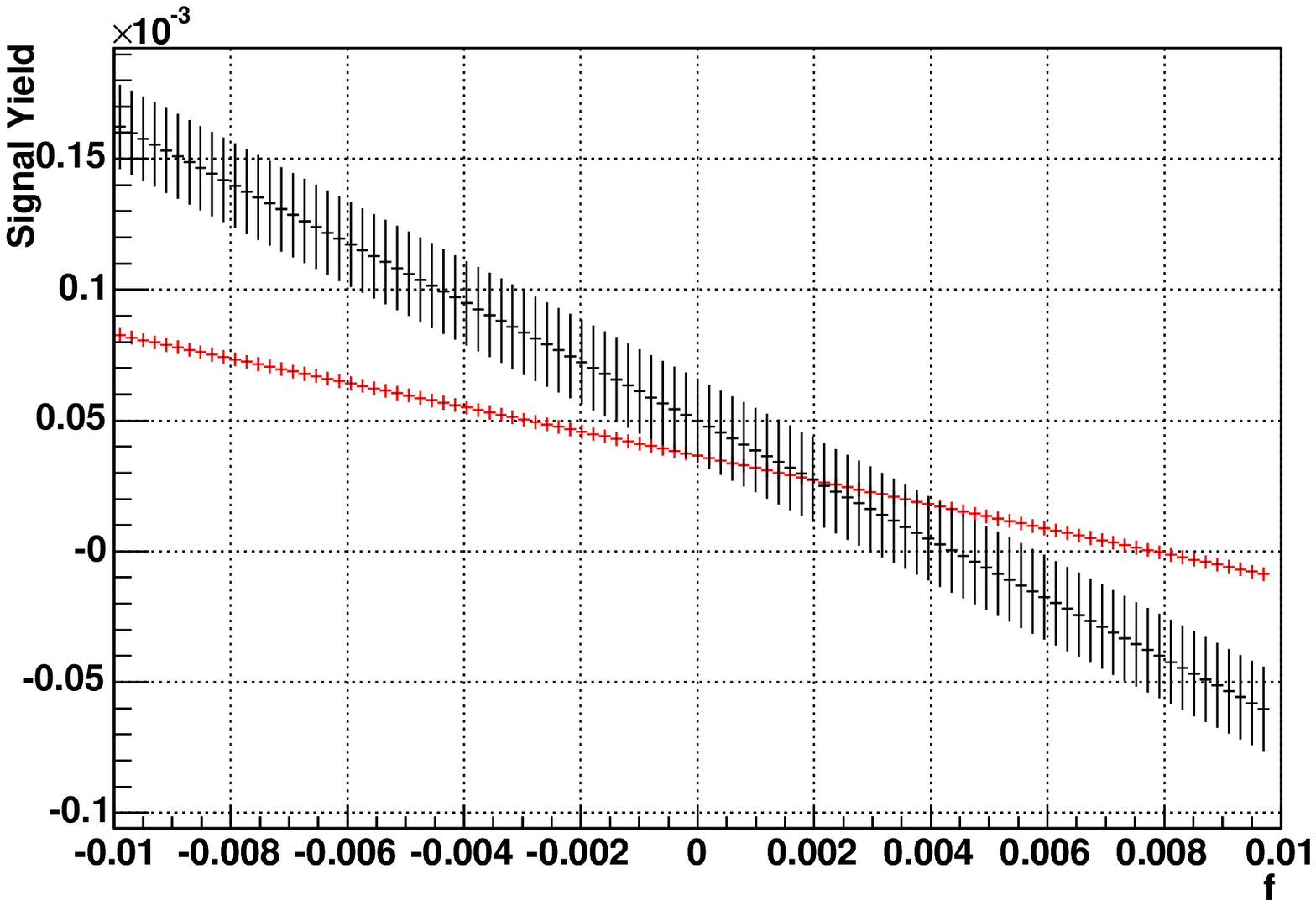,width=0.33\textwidth}}}}
 \caption{\label{fig:errnorm} Signal yield $S$ for converter (black) and non converter (red) runs as a function of the possible background normalization error $f$.}
\end{figure}
As shown in Fig.~\ref{fig:mass}, the results 
from both data sets agree well within statistical errors, which are 30\% in 
the range from 150 to 750 MeV/c$^2$ and less for lower masses. Considering the 
decreased $S/B$ ratio for the data with the converter we can estimate a 0.1\% 
scale uncertainty of the background normalization, well within the 0.25\% 
systematic uncertainty assigned.
Indeed if we were to make a mistake in the background subtraction by a factor $f$, we get different signal in both cases:
\begin{eqnarray}
S_{non converter} = FG - (1+f) \cdot BG = S - f BG \\
S_{converter} = FG_{converter} - 2.5 \cdot (1+f) \cdot BG = S - 2.5 f BG
\end{eqnarray}
We measured that the agreement between converter and non converter signal $b=S_{non converter}$/$S_{converter} \leq 1.3$. Therefore
\begin{eqnarray}
b= \frac{ S_{non converter} } { S_{converter} } = \frac{(S -f BG)} { (S - 2.5 f BG)} \\
b S - 2.5 b f BG = S - f BG
\end{eqnarray}
and solving for $f$ we obtain
\begin{eqnarray}
(b-1) S = (2.5 b - 1) \cdot f BG \\
f = \frac{(b-1)}{(2.5 b-1)} \cdot \frac{S}{BG}
\end{eqnarray}
In our case $f = 0.3/2.25 * S/BG = 0.13 \cdot S/B$ 
with $S/B  \sim 1/150$ as in the low mass region, we finally can constrain $f = 0.001$ or 0.1\%. 
Figure \ref{fig:errnorm} shows the signal yield $S$ for converter (black) and non converter (red) runs as a function of the possible background normalization error $f$.
The signal yield for converter and non converter runs coincides at small or null values of $f$, verifying that the background normalization is estimated correctly within the uncertainties quoted.

%%%%%%%%%%%%%%%%%%%%%%%%%%%%%%%%%%%%%%---efficiency correction

\subsection{Efficiency Correction}\label{ssec:effi}
The spectra are corrected for electron detection efficiency, to give the e$^+$e$^-$--pair yield for both e$^+$ and e$^-$ inside the PHENIX aperture,
as specified in Eq.~\ref{eq:acc}. The correction is determined using a GEANT 
simulation \cite{geant} of the PHENIX detector that includes the details of 
the detector response. 
\begin{figure*}[t]
 \includegraphics[width=0.625\linewidth]{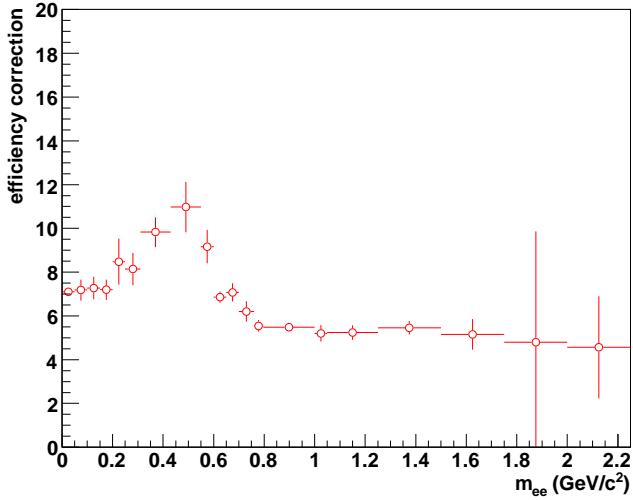}
 \caption{  \label{fig:effi}
 Effective efficiency correction as a function of invariant mass, obtained by weighting the (m, p$_\mathrm{T}$) map with the p$_\mathrm{T}$ distribution of the minimum bias data set.}  
\end{figure*}
About 1 million e$^+$e$^-$--pairs are generated: about half of them were generated flat in mass (from 0 to 4\,GeV), p$_\mathrm{T}$ (from 0 to
4\,GeV), azimuthal angle (from 0 to $2\pi$), in 1 unit of rapidity.
In the other half the initial mass and p$_\mathrm{T}$ distribution have been chosen linearly falling
to enhance the statistics in the low mass and p$_\mathrm{T}$ region, where the efficiency varies most.
Only pairs with both electrons and positron in the acceptance (\ref{eq:acc}) are processed by GEANT, reconstructed with the same analysis chain and all cuts applied. The correction is determined double differentially in p$_\mathrm{T}$ and mass of the e$^+$e$^-$--pair.
The efficiency has been determined and applied double differentially in mass and p$_\mathrm{T}$; 
the same efficiency correction (mass, p$_\mathrm{T}$) is applied to all centrality bins.
Although this procedure avoids an assumption for the input kinematic distribution, whenever the statistics of the data is limited (as for example when correcting the individual centralities), it may suffer statistical fluctuations.
In this case we use also an effective efficiency correction, as a function of mass only, obtained by weighing the 2-dimensional (mass, p$_\mathrm{T}$) corrections with a realistic p$_\mathrm{T}$ distribution provided by the minimum bias data set, as shown in Figure \ref{fig:effi}. We take the difference of 10\% between the 1D and the 2D corrected result as additional systematic on the efficiency correction of the different centrality data sets ( see \ref{fig:ratio}).  
Figure \ref{fig:effi} shows the effective efficiency correction, weighted with the p$_\mathrm{T}$ distribution of the minimum bias data set. 
At high masses the efficiency is constant; its value is essentially the square of the single electron efficiency ($1/5.7=\epsilon_{pair}\approx\epsilon_{single}^2$) which is $\sim$ 40\% and depends on the track reconstrction (90\%), the electron identification (65\%) and the dead areas (70\%) within the acceptance.
At low masses the efficiency results from the convolution of the single reconstruction efficiency which drops toward low momenta and the acceptance which effectively truncates the single p$_\mathrm{T}$ distribution. From 800 to 400 MeV/c$^2$ the pair efficiency drops as a consequence of the drop at low p$_\mathrm{T}$ of the single electron efficiency. However, below 400 MeV/c$^2$, the acceptance hole (which cuts single tracks with $p_T\leq 200 MeV/c$) truncates the single particle distribution at lower p$_\mathrm{T}$ leading to a larger average momentum; and the efficiency consequently increases.
The reduction of the electron reconstruction efficiency (0.92$\pm$0.03) due to detector occupancy is corrected.

%%%%%%%%%%%%%%%%%%%%%%%%%%%%%%%%%%%%%%---systematic error
The systematic uncertainties can be summarized as: (i) 13.4\% 
on dielectron reconstruction, which is twice the uncertainty on the electron 
reconstruction efficiency \cite{ppg066}, (ii) 6\% conversion rejection cut, 
(iii) 5\% event rejection and (iv) 3\% occupancy. These uncertainties are 
included in the final systematic error on the invariant e$^+$e$^-$--pair 
yield. For the individual centrality bins, we added 10\% uncertainty arising from the p$_\mathrm{T}$ of the efficiency correction.

%%%%%%%%%%%%%%%%%%%%%%%%%%%%%%%%%%%%%%---fig 2
%%%%%%%%%%%%%%%%%%%%%%%%%%%%%%%%%%%%%%---cocktail
\subsection{Cocktail of Hadronic Sources and Charmed Mesons}\label{ssec:cocktail}

\begin{figure*}[t]
 \includegraphics[width=0.625\linewidth]{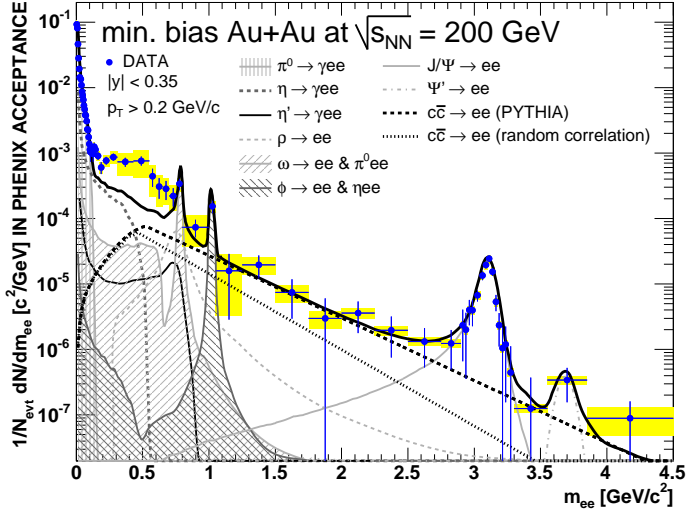}
 \caption{  \label{fig:cock}
Invariant e$^+$e$^-$--pair yield compared to the yield from the model of 
hadron decays. The charmed meson decay contribution based on PYTHIA is 
included in the sum of sources (solid black line). The charm contribution 
expected if the dynamic correlation of $c$ and $\bar{c}$ is removed is shown 
separately. Statistical (bars) and systematic (boxes) uncertainties are shown 
separately; the mass range covered by each data point is given by horizontal 
bars. The systematic uncertainty on the cocktail is not shown.}
\end{figure*}

Figure~\ref{fig:cock}\cite{ppg075} compares the invariant yield to the expected yield from 
meson decays and correlated decays of charmed mesons. The cocktail of hadron 
decay contributions was estimated using PHENIX data for meson production when 
available.  As input distributions we use the measured $\pi, \eta, \phi, 
J/\psi$ yield and spectra \cite{pich, pi0, eta, ppg016, ppg068}.  For other 
mesons we use the $m_{\mathrm{T}}$ scaling procedure outlined in~\cite{ppg066}. 
The systematic uncertainties depend on mass and range from 10 to 25\% \cite{ppg066}. They result from the uncertainty on the measured pion yield and on all meson-to-pion ratios.

%%%%%%%%%%%%%%%%%%%%%%%%%%%%%%%%%%%%%%---charm

For the continuum below the $J/\psi$ the dynamic correlation of $c$ and 
$\bar{c}$ is essential, but modified to an unknown extent in Au+Au collisions compared to p+p collisions. We make two assumptions: (i) the 
correlation is unchanged by the medium and equal to what is known from p+p 
collisions. In this case we can use PYTHIA \cite{pythia} scaled from the p+p 
equivalent $c\overline{c}$ cross section of 567$\pm$57$\pm$193 $\mu$barn 
\cite{ppg065} to minimum bias Au+Au collisions proportional to the mean number of 
binary collisions ($258 \pm 25$) \cite{pich}. We note that the p$_\mathrm{T}$ 
distribution for electrons generated by PYTHIA is softer than the spectra 
measured in p+p data but coincides with those observed in Au+Au 
\cite{ppg066}. As a second assumption (ii) there is no dynamical correlation, 
i.e. the direction of $c$ and $\bar{c}$ quarks are uncorrelated. We keep the 
overall cross section and the p$_\mathrm{T}$ distributions fixed to 
experimental data \cite{ppg066}. Other contributions from bottom and 
Drell-Yan are expected to be small in the mass region below the $J/\psi$ 
peak. Each e$^+$ and e$^-$ must fall in the PHENIX acceptance, given by 
Eq.~\ref{eq:acc}.

%%%%%%%%%%%%%%%%%%%%%%%%%%%%%%%%%%%%%%---description of fig2

The data below 150 MeV/c$^2$ are well described by the cocktail of 
hadronic sources. The vector mesons $\omega$, $\phi$ and $J/\psi$ are 
reproduced within the uncertainties. However, the yield is substantially 
enhanced above the expected yield in the continuum region from 150 to 750 
MeV/c$^2$. The enhancement in this mass range is a factor of 
3.4$\pm$0.2(stat)$\pm$1.3(syst)$\pm$0.7(model), where the first error is 
the statistical error, the second the systematic uncertainty of the data, 
and the last error is an estimate of the uncertainty of the expected 
yield. 
Note that the last two errors are conservatively estimated and are not 1 $\sigma$ equivalent.
Above the $\phi$ meson mass the data seem to be well described by 
the continuum calculation based on PYTHIA. This is somewhat surprising, 
since single electron distributions from charm show substantial medium 
modifications \cite{ppg066}, and thus it is hard to understand how the 
dynamic correlation at production of the $c\bar{c}$ remains unaffected by 
the medium. A complete randomization of that correlation (see 
Fig.\ref{fig:cock}) leads to a much softer mass spectrum and would leave 
significant room for other contributions, e.g. thermal radiation.

%%%%%%%%%%%%%%%%%%%%%%%%%%%%%%%%%%%%%%---fig : p+p
\subsection{Comparison with p+p Data}
We also analyzed a set of 65 million p+p collisions collected by PHENIX during the 2005 RHIC run. The data were recorded using a single electron trigger in addition to the reaction trigger with the BBC. The electron trigger is defined by a matching between the EMCal and RICH in a small angular area with a minimum energy deposition of 0.4 GeV in any 2x2 patch of EMCal towers. 
The p+p analysis is done similar to the Au+Au analysis discussed above, but with less strict electron identification cuts.
\begin{figure}[!h]
\mbox{
\parbox{8cm}{\subfigure[]{\epsfig{file=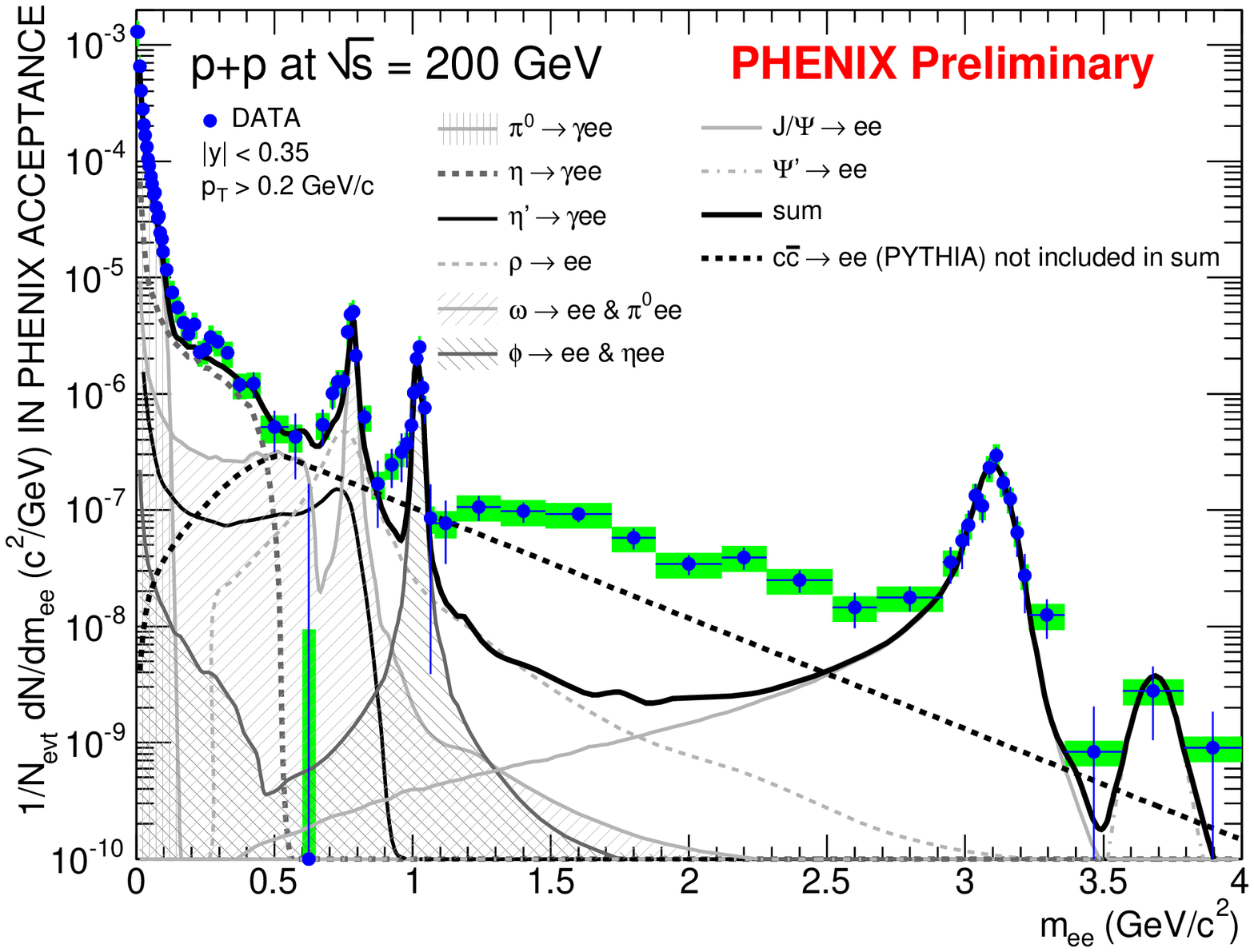,width=0.5\textwidth}}}
\parbox{8cm}{\subfigure[]{\epsfig{file=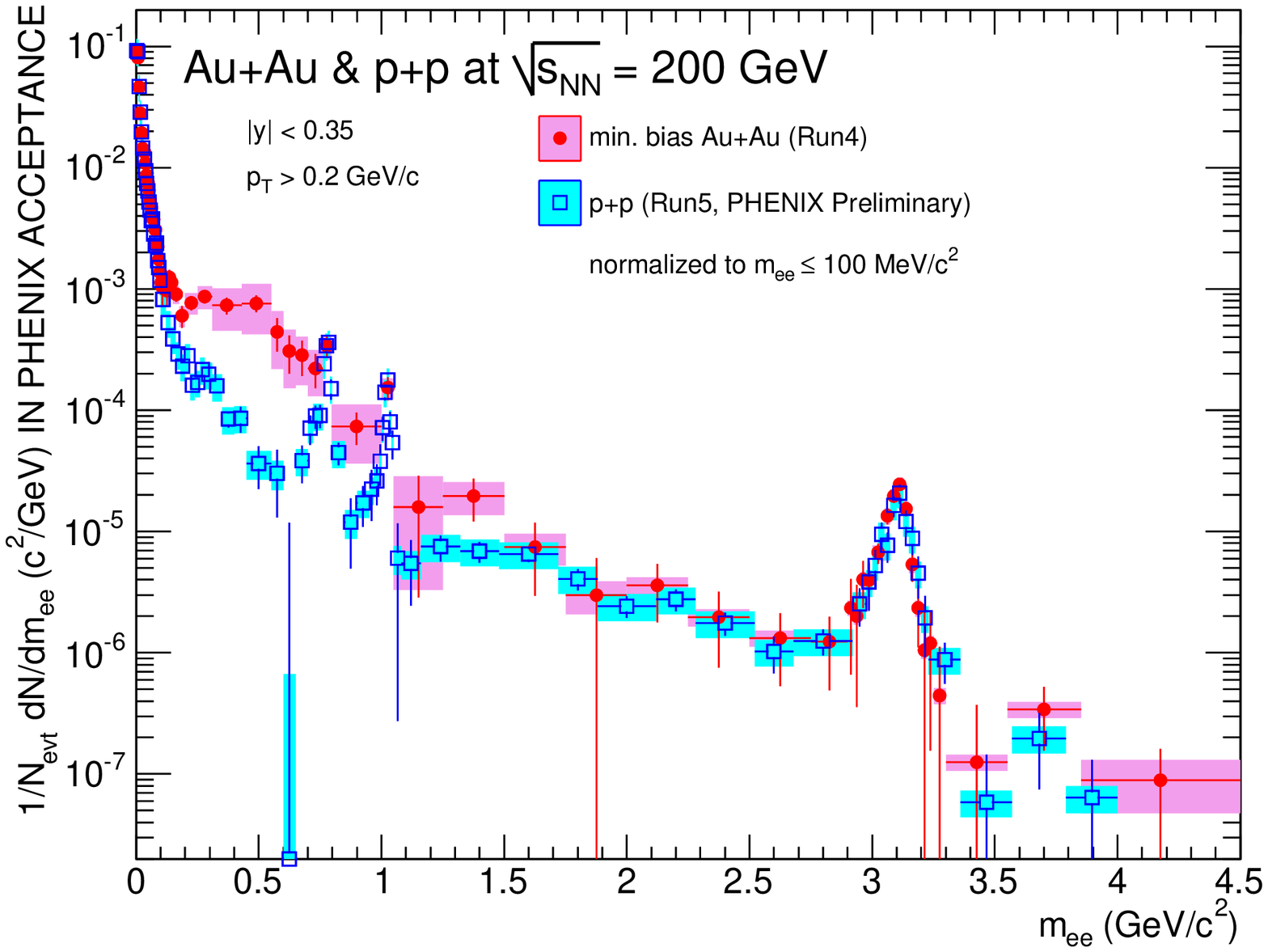,width=0.5\textwidth}}}}
 \caption{\label{fig:pp}
   Invariant e$^+$e$^-$--pair yield in p+p collisions compared to the yield from the model of 
   hadron decays. The charmed meson decay contribution based on PYTHIA is not
   included in the sum of sources (solid black line). In the right panel the e$^+$e$^-$--pair yield measured in p+p is compared to the yield measured in Au+Au. 
   Statistical (bars) and systematic (boxes) uncertainties are shown separately; the mass range covered by each data point is given by horizontal bars.}
\end{figure}
The single electron trigger increases the average p$_\mathrm{T}$ of the electrons in the sample, and therefore modifies the shape of the combinatorial
background. 
To start of the correct single electron p$_\mathrm{T}$ distributions the mixed events have been generated from the
Minimum Bias data sample. The normalization of the combinatorial background is done in the same way as in the Au+Au analysis but the limited statistics in the like-sign pairs results in an increase of the statistical error on the background normalization to 2\% This does not reduce the significance of the signal, because of the higher signal-to-background ratio. Correlated background from conversion and cross pairs has been subtracted as in the Au+Au analysis. The data have been corrected for the reconstruction efficiency and for the trigger efficiency. The trigger efficiency has been determined implementing in a fast Monte Carlo of hadron decays a parameterization of the trigger efficiency as function of the single electron p$_\mathrm{T}$ for every EMCal sector. 
Finally the data has been scaled such that the J/$\psi$ yield corresponds to the yield published in \cite{ppg069} but filtered in the PHENIX acceptance.
Figure~\ref{fig:pp} (left panel) compares the invariant yield to the expected yield from 
meson decays calculated as in \ref{ssec:cocktail} but using a parameterization of $\pi^0$ and $\pi^{\pm}$ measured in p+p \cite{pi0_pp, pich_pp}. The same PYTHIA calculation used in Au+Au for the charmed mesons is shown in the figure but not added to the cocktail.
The agreement with the cocktail is very good in the low invariant mass range, the resonances are well reproduced. In the intermediate mass region the data are not well described by PYTHIA; this could be explained by the disagreement between the p$_\mathrm{T}$ 
distribution for electrons generated by PYTHIA and the spectra measured in p+p and might have consequences even in the low mass region where PYTHIA predicts a substantial $e^+e^-$ yield. Also jet-induced correlations could have a contribution in this mass region. 
In figure~\ref{fig:pp} (right panel) we compare the invariant yield measured in p+p with the yield measured in Au+Au, scaled to the integral in the region $m_{ee}\leq 100 MeV/c^2$. While the $\omega$ and $\phi$ resonances are consistent within the uncertainties, the continuum yield in the low mass region is significantly enhanced in the Au+Au data with respect to the p+p, confirming the enhancement measured with respect to the hadronic cocktail. In the intermediate mass region, above 1.5 GeV/c$^2$ the spectra agree. However, in this region the Au+Au data may be the result of two competing effects, i.e. a 'softening' of the charm spectrum (as indicated by the dashed line of figure \ref{fig:cock}) due to energy loss of the charmed mesons in the medium and a thermal source. The agreement of the two spectra continues in the J/$\psi$ peaks, indicating a similar scaling for the J/$\psi$ as for the pions, or that the measured J/$\psi$ suppression \cite{ppg069} is 'compensated' by the relative increase of $N_{coll}$/$N_{part}$.    
\begin{figure}[!h]
 \includegraphics[width=0.7\linewidth]{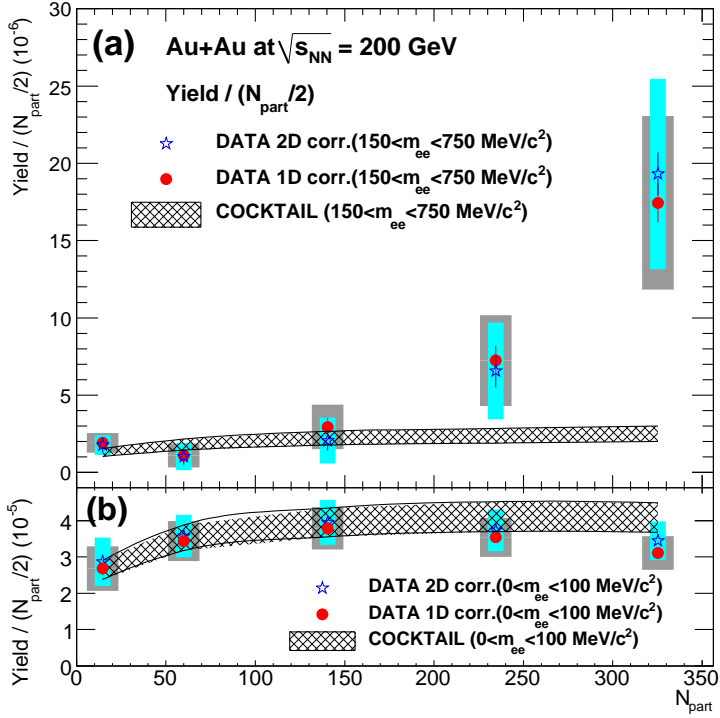}
 \caption{\label{fig:ratio}
Dielectron yield per participating nucleon pairs ($N_{\rm part}/2$) as 
function of $N_{part}$ for two different mass ranges compared to the expected 
yield from the hadron decay model. The two lines give the 
systematic uncertainties. For the data statistical and systematic uncertainties 
are shown separately.}
\end{figure}

%%%%%%%%%%%%%%%%%%%%%%%%%%%%%%%%%%%%%%---fig 3
\subsection{Centrality Dependency}
To shed more light on the continuum yield we have studied the centrality 
dependence of the yield in three mass windows, below 100 MeV/c$^2$, from 
150 to 750 MeV/c$^2$ and 1.2 to 2.8 GeV/c$^2$. 
Using simulations based on a Glauber model calculation \cite{glau} the average number of participants $N_{part}$ and the average number of binary collisions $N_{coll}$ associated with each centrality bin is determined.
The centrality bins cover 0-10\%, 10-20\%, 20-40\%. 40-60\%, 60-92\% fraction of the Au+Au minimum bias cross section.
The top panel of Fig.~\ref{fig:ratio} \cite{ppg075} shows the centrality dependence of the yield in the 
mass region 150--750 MeV/c$^2$ divided by the number of participating 
nucleon pairs ($N_{\rm part}/2$). For comparison the yield below 100 
MeV/c$^2$, which is dominated by low p$_\mathrm{T}$ pion decays, is shown 
in the lower panel. For both intervals the yield is compared to the same 
yield calculated from the hadron cocktail. In the lower mass range the 
yield agrees with the expectation, i.e. is proportional to the pion yield. 
In contrast, in the range from 150 to 750 MeV/c$^2$, the observed yield 
rises significantly compared to the expectation, reaching a factor of 
6.6$\pm$0.5(stat)$\pm$2.1(syst)$\pm$1.3(model) for most central 
collisions. The increase is qualitatively consistent with the conjecture 
that an in-medium enhancement of the dielectron continuum yield arises 
from scattering processes like $\pi\pi$ or $q\bar{q}$ annihilation, which 
would result in a yield rising faster than proportional to $N_{part}$. 
The different sets of data in the figure shows the difference (discussed in \ref{ssec:effi}) between the 1D and 2D efficiency correction which is well within the quoted systematic uncertainties.

%%%%%%%%%%%%%%%%%%%%%%%%%%%%%%%%%%%%%%---fig 4

\begin{figure}[t]
 \includegraphics[width=0.7\linewidth]{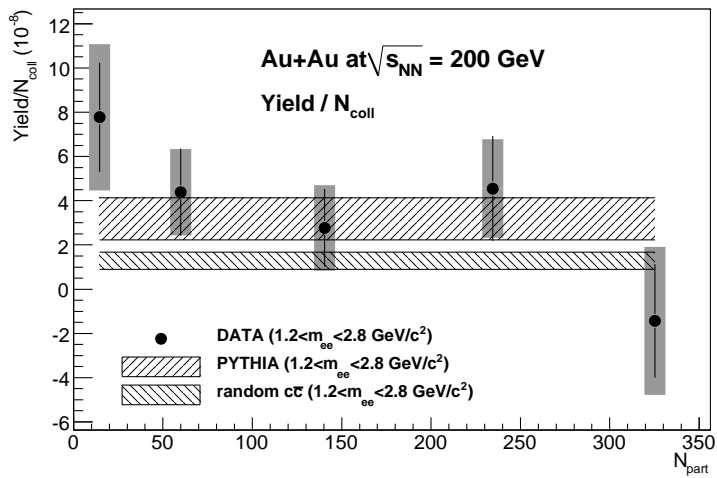}
 \caption{  \label{fig:ratio2}
Dielectron yield per number of collisions $N_{coll}$ in the mass range 1.2 to 
2.8 GeV/c$^2$ as function of $N_{\rm part}$. Statistical and systematic 
errors are shown separately. Also shown are two bands corresponding to the 
two different estimates of the contribution from charmed meson decays. The 
width of the band reflect the uncertainty of the charm cross-section only.}
\end{figure}

We normalize the yield in the mass region 1.2 to 2.8 GeV/c$^2$ to the number 
of binary collisions (Fig.~\ref{fig:ratio2} \cite{ppg075}), which is the correct scaling 
for pairs from charmed meson decays \cite{ppg066}. The normalized yield shows 
no significant centrality dependence and is consistent with the expectation 
based on PYTHIA. It is also likely that a scenario where the correlation 
between the $c$ and $\bar{c}$ is randomized will require an additional 
source, e.g. a contribution from thermal radiation. This contribution could 
increase faster than linearly with $N_{\rm part}$ and therefore the apparent 
scaling with $N_{coll}$ may be a mere coincidence. We note that such a
coincidence may have been observed in this mass region at the CERN SPS 
\cite{NA50}, where a major prompt component has now been suggested by NA60 
data \cite{NA60_therm}.

\section{Conclusions}
%%%%%%%%%%%%%%%%%%%%%%%%%%%%%%%%%%%%%%--- conclusions 

In conclusion, measurements of Au+Au collisions at $\sqrt{s_{NN}}$=200 GeV show  an enhancement of the dielectron continuum in the mass range 150--750 MeV/c$^2$ in central collisions. The enhancement is absent in peripheral collisions, where the observed yield agrees well with the calculated hadronic background.
Enhancement in similar mass regions was reported by lower energy experiments \cite{CER2, NA60_rho, HADES}.
The observed yield between $\phi$ and $J/\psi$ is consistent with
the expectation from correlated $c\bar{c}$ production, but does not exclude
other mechanisms.

\end{document}